\documentclass[english,prl,twocolumn,footinbib]{revtex4}
\usepackage[T1]{fontenc}
\usepackage[latin1]{inputenc}
\usepackage{graphicx}
\usepackage{amssymb}

\makeatletter

\providecommand{\LyX}{L\kern-.1667em\lower.25em\hbox{Y}\kern-.125emX\@}

\usepackage{babel}
\makeatother
\begin{document}

\title{Electron in a tangled chain: multifractality at the small-world critical
point}

\begin{abstract}
We study a simple model of conducting polymers in which a single electron
propagates through a randomly tangled chain. The model has the geometry
of a small-world network, with a small density $p$ of crossings of
the chain acting as shortcuts for the electron. We use numerical diagonalisation
and simple analytical arguments to discuss the density of states,
inverse participation ratios and wave functions. We suggest that there
is a critical point at $p=0$ and demonstrate finite-size scaling
of the energy and wave functions at the lower band edge. The wave
functions are multifractal. The critical exponent of the correlation
length is consistent with criticality due to the small-world effect,
as distinct from the previously discussed, dimensionality-driven Anderson
transition.
\end{abstract}

\author{J. Quintanilla}

\affiliation{
School of Physics \& Astronomy, University of Birmingham, Edgbaston,
Birmingham, West Midlands, B15 2TT, U.K.
\\
ISIS facility, CCLRC Rutherford Appleton Laboratory, Chilton, 
Didcot, Oxfordshire, OX11 0QX, U.K.
}

\author{V. L. Campo}

\affiliation{
Departamento de F\'{\i}sica e Inform\'atica, 
Instituto de F\'{\i}sica de S\~ao Carlos, 
Universidade de S\~ao Paulo, Caixa Postal 369,
13560-970 S\~ao Carlos, S\~ao Paulo, Brazil. 
\\
Centro Internacional de F\'{\i}sica da Mat\'{e}ria Condensada,
Universidade de Bras\'{\i}lia, Caixa Postal 04513, 70910-970, Bras\'{\i}lia, 
Brazil.
}

\maketitle

\section*{Introduction}

The small-world effect occurs in complex networks \cite{Albert-Barabasi-02}
that resemble regular lattices with a small density $p$ of ``shortcuts''
per site \cite{Watts-Strogatz-98}. The key observable is the shortest
path length between two randomly chosen points, $l$. Defining a ``thermodynamic
limit'' by \begin{equation}
L\to \infty \, ,\, p=\textrm{constant},\label{eq:thermodynamic-limit}\end{equation}
 where $L$ is the number of sites, one finds that for $0<p\ll 1$ the
network has the local structure of a regular lattice but some global
properties of a random graph \cite{Watts-Strogatz-98}. In fact there
is a \emph{critical point} \cite{Newman-Watts-99,Newman-Watts-99b}
at $p=0$ %
It separates regular lattices ($p=0$), for which the relative size
of the shortest path length, $l/L$, tends to some finite value, from
lattices with finite density of shortcuts ($p>0$), for which $l/L\to 0$.
This quantity shows finite-size scaling \cite{Newman-Watts-99,Newman-Watts-99b,Barthelemey-Amaral-99,Barrat-Weigt-00}:
for \begin{equation}
L\gg 1,\, p\ll 1\label{eq:finite-size-scaling}\end{equation}
it depends only on the scaling variable $L/\xi $, where $\xi =p^{-\nu }$
is a critical system size diverging algebraically at $p=0$. A simple
renormalisation group argument yields the critical exponent $\nu $
\cite{Newman-Watts-99}.

On the other hand it has been accepted for some time that conducting
polymers must be thought of as complex networks \cite{Prigodin-Efetov-93}.
Models have been studied in which a one-dimensional Anderson insulator
is perturbed by adding some additional connectivity in the form of
a given density of chain {}``crossings''. It is well established
that, when the density of crossings reaches a finite critical value,
which depends on the amount of site-energy disorder, there is an Anderson
metal-insulator transition \cite{Prigodin-Efetov-93,Zambetakiy-Evangelou-Economou-96,Dupuis-97}.
This critical point results from the increase of the effective dimensionality
of the network. It has also been demonstrated \cite{Zhu-Xiong-00,Zhu-Xiong-01}
in a model having the geometry of the Watts-Strogatz (WS) small-world
network \cite{Watts-Strogatz-98}. 

Interestingly, even models of conducting polymers that lack site-energy
disorder, so that the state of a single wire with no crossings is
metallic, display non-trivial behaviour suggestive of proximity to
a second-order phase transition \cite{Xiong-Evangelou-95,Xiong-Chen-Evangelou-96}.
But there has been no detailed discussion of the nature of this critical
point or its location on the phase diagram. Here we shall provide
evidence that it occurs when the density of crossings reaches zero
and that its origin is the small-world effect. 

The rest of paper is organised as follows. First we introduce
a simple \emph{model} that generalises the {}``winding chain'' of
Ref.~\cite{Xiong-Evangelou-95} and has the geometry of the Newman-Watts
(NW) network \cite{Newman-Watts-99}, with the crossings acting as
shortcuts. Then we present results of full-spectrum numerical diagonalisation
that demonstrate the \emph{small-world effect}: a qualitative change
of the spectrum and wave functions on the addition of a small density
of shortcuts. We then use Lanczos diagonalisation to study in detail,
for much larger systems, the \emph{lower band edge}. We find finite-size
scaling consistent with the critical behaviour of the network described
in \cite{Newman-Watts-99,Newman-Watts-99b}. Simple analytical arguments
are used to characterise the wave functions at the critical point.
Our conclusions are then laid out.

\section*{Model} We consider a single particle moving in a tight-binding
lattice with one orbital per site. The site energy is equal on all
$L$ sites. Hopping can take place between nearest neighbours or between
sites connected by one of $N$ shortcuts. For simplicity we assume
the same value of the hopping integral, $t$. The Hamiltonian is\begin{equation}
\hat{H}=-t\sum _{j}\left|j\right\rangle \left\langle j+1\right|-t\sum _{s=1}^{N}\left|i_{s}\right\rangle \left\langle j_{s}\right|+\textrm{H}.\textrm{c}.\label{eq:hamiltonian}\end{equation}
where $j$ is a site label and $s$ denotes the shortcut connecting
sites $i_{s}$ and $j_{s}$. We assume periodic boundary conditions:
$\left|L+1\right\rangle =\left|1\right\rangle $. 

Note that $t$ is the only energy scale in the Hamiltonian (\ref{eq:hamiltonian}):
the system's behaviour is thus completely specified by the way in
which the $N$ shortcuts are chosen. One can imagine creating these
shortcuts by winding an initially straight chain, embedded in a higher-dimensional
space, in a random fashion. Sites that were initially far apart may
come close so tunnelling of the electron between these sites becomes
possible. 

The model of conducting polymers in Ref.~\cite{Xiong-Evangelou-95}
is obtained by requiring that such winding is in the form of a series
of consecutive, non-intersecting loops: $i_{1}<j_{1}<i_{2}<j_{2}<\ldots <i_{N}<j_{N}$
(except for the boundary conditions, which do not affect our argument).
Here we relax that restriction: our chain is {}``tangled'' at random
so that each of 
\begin{equation}
\mathcal{N}\equiv \frac{L\left(L-3\right)}{2} \label{eqN}
\end{equation}
possible shortcuts (excluding only shortcuts joining a site to itself or to
a nearest neighbour) is present with the same probability, 
\begin{equation}
\mathcal{P}\equiv \frac{2p}{\left(L-3\right)}. \label{eqP}
\end{equation}
The resulting topology is an instance of the NW network with critical
exponent $\nu =1$ %
\footnote{One can envisage more realistic generalisations of the model in \cite{Xiong-Evangelou-95}
that take into account correlations between the positions of different
crossings \cite{Luque-Miramontes-02,Metzler-et-al-02} or their unquenched
nature \cite{Bao-Lee-Quake-03}. Our choice of model is dictated by
simplicity and by analogy with the Anderson model (which features
quenched, uncorrelated site-energy disorder). %
}.

A similar Hamiltonian, namely the adjacency matrix of the WS network,
has been considered in \cite{Farkas-Derenyi-Barabasi-Vicsek-01,Kim-Hong-Choi-03}.
The authors of \cite{Kim-Hong-Choi-03} solved the time-dependent
Schr\"odinger equation, demonstrating the faster spread of initially
localised wave functions on a small-world network compared to a regular
lattice. In \cite{Farkas-Derenyi-Barabasi-Vicsek-01} the eigenvalues
and eigenvectors were obtained, as a means of characterising the networks'
geometry. Our argument is based on a similar calculation. Note, however,
that our model has a distinct topology, compatible with the simple
picture of a tangled chain---indeed the Hamiltonian of \cite{Farkas-Derenyi-Barabasi-Vicsek-01,Kim-Hong-Choi-03}
is not of the form (\ref{eq:hamiltonian}). Moreover our specific
aim is to identify critical behaviour near $p=0$ %
\footnote{A related class of models, describing vibrations of tangled chains,
results when the Laplace operator is defined on the NW \cite{Monasson-99}
or WS networks \cite{Kim-Hong-Choi-03b}.%
}. 

In our model the number of shortcuts, $N$, is a random variable following
the binomial distribution (Poisson for $L\gg p$). Evidently $\left\langle N\right\rangle =Lp$,
where $\left\langle \ldots \right\rangle $ denotes a configurational
average. 

\section*{Small-world effect} For a finite size $L$ and a given set
of shortcuts (a {}``realisation'') it is straightforward to diagonalise
the Hamiltonian (\ref{eq:hamiltonian}) numerically %
\footnote{We used the diagonalisation function of \cite{Octave}. %
}. One can then, by averaging over a finite number of realisations,
$R$, obtain a coarse-grained estimate of the configurationally averaged
density of states (DOS), $\rho \left(\epsilon \right)\equiv \left\langle \sum _{\nu }\delta \left(\epsilon -\epsilon _{\nu }\right)\right\rangle ,$
for a particular value of $p$. Here $\epsilon _{\nu }$ is the energy
of the $\nu ^{{\mbox {\small th}}}$ state.

\begin{figure}
\includegraphics[  width=0.90\columnwidth,
  keepaspectratio]{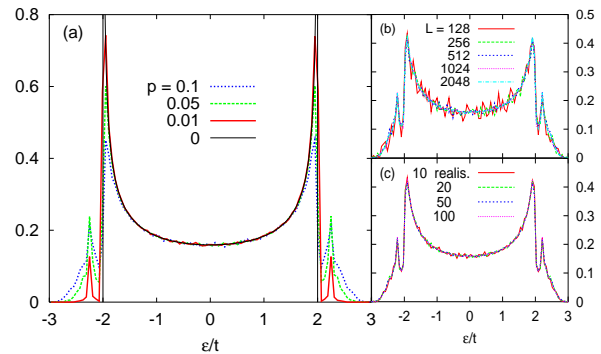}

\caption{\label{cap:DOS}(Color online) Density of states (DOS), estimated from full-spectrum
numerical diagonalisation for (a) $R=32$ realisations, $L=1024$
sites and different densities of shortcuts per site, $p$; (b) $R=16$,
$p=0.1$ and different values of $L$; (c) different $R$ with $L=1024$
and $p=0.1$. All the plots have been obtained by drawing histograms
using 100 uniformly spaced values of the energy from $\epsilon =-3t$
to $\epsilon =+3t$. }
\end{figure}

The main panel of Fig.~\ref{cap:DOS} shows the DOS, estimated in
this way, for several small values of $p$. The curve for $p=0$,
given by the well-known formula $\rho \left(\epsilon \right)=\left(2\pi \sqrt{1-\epsilon ^{2}/4t^{2}}\right)^{-1},$
is also shown. Panels (b) and (c) illustrate the dependences of the
numerical results on $L$ and on $R$. They suggest that the DOS is
well defined in the thermodynamic limit (\ref{eq:thermodynamic-limit}),
and that the plot on the main panel is representative of it, at this
level of coarse-graining (higher values of $L$ and $R$ would be
required to describe the finer detail). At $p=0$, the DOS has the
usual square-root singularities at the band edges. For $p\gtrsim 0$,
the singularities survive in the form of pronounced maxima in the
DOS. This is similar to what was described for the model based on
the WS network \cite{Farkas-Derenyi-Barabasi-Vicsek-01}. Also in
common with that model, for $p\gtrsim 1$ (not shown), the spectral
density resembles that obtained by numerical diagonalisation of the
adjacency matrices of uncorrelated random graphs. However the salient
features of Fig.~\ref{cap:DOS} are two additional peaks outside
the conduction band of the chain, located at $\epsilon =\pm \sqrt{5}t$.
Although this region of the spectrum becomes less populated as $p\to 0$
the position of the peaks does not change and they are present for
any non-zero value of $p$. They are a qualitative feature distinguishing
the DOS for $p=0$ from that for $p\gtrsim 0$. 

\begin{figure}
\includegraphics[  width=1.0\columnwidth,
  keepaspectratio]{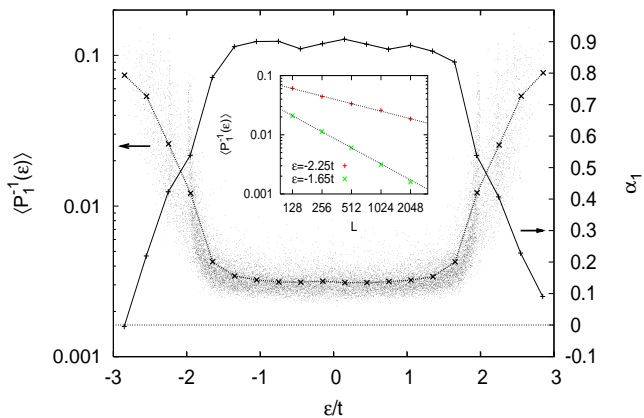}

\caption{\label{cap:IPR-vs-energy}(Color online) Energy-dependence of the IPR (left-hand
axis) and the exponent $\alpha _{1}$ characterising its power-law
dependence on system size $L$ (right-hand axis). The data have been
obtained by  exact numerical diagonalisation for $R=16$ realisations
with $p=0.1$ shortcuts/site. The IPR vs energy curve corresponds
to $L=1024$ sites. It has been obtained by splitting the energy interval
from $\epsilon =-3t$ to $\epsilon =+3t$ into 20 sub-intervals and
averaging the IPR over each subinterval. The 16384 individual values
of the IPR in our ensemble are also shown. Similar curves have been
obtained for different values of $L$, and fits of the averaged IPR
to a power-law in each energy interval have been used to determine
$\alpha _{1}$. Two examples of such fits are shown in the inset.}
\end{figure}

To explore further the nature of the states around these new peaks
we can use numerical diagonalisation to obtain the inverse participation
ratio (IPR; $n=1$) and higher-order moments of the local density
distribution ($n=2,3,\ldots $). These are given by $P_{n}^{-1}\left(\epsilon \right)\equiv \sum _{j\nu }\delta \left(\epsilon -\epsilon _{\nu }\right)\left|\psi _{\nu }\left(j\right)\right|^{2\left(n+1\right)}$
where $\psi _{\nu }\left(j\right)$ is the wave function of the $\nu ^{\mbox {\small th}}$
state, evaluated on the $j^{\mbox {\small th}}$ site. For large enough
values of $L$, we expect the configurationally-averaged IPR to be
given by a power law: $\left\langle P_{1}^{-1}\left(\epsilon \right)\right\rangle =L^{-\alpha _{1}}.$
The coefficient $\alpha _{1}$ can be used to distinguish extended
states ($\alpha _{1}=1$) from localised states ($\alpha _{1}=0$)
and states with fractal dimension ($0<\alpha _{1}<1$) \cite{Kramer-MacKinnon-93}.
Fig.~\ref{cap:IPR-vs-energy} shows our results. They suggest that,
for $p=0.1$, all states are fractal, but the wave functions with
$\left|\epsilon \right|>2t$ are much closer to localisation than
those with $\left|\epsilon \right|<2t$, which are almost extended.
Note the strong fluctuations of the IPR at particular energies. 

\begin{figure}
\includegraphics[  width=0.75\columnwidth,
  keepaspectratio]{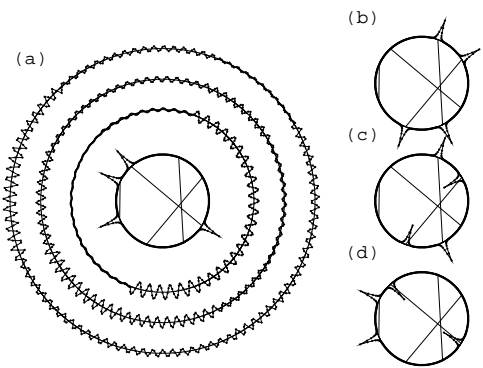}

\caption{\label{cap:wave-functions-polar}Numerically determined single-particle
wave functions for a particular realisation of our tight-binding model,
with $L=460$ sites. Each wave function has been represented on a
circle, whose radius is its energy, measured from $-3t$. The innermost
wave function in (a) is the ground state. The three excited states
with energy below $-2t$ have been plotted separately for clarity:
(b), (c) and (d), in order of increasing energy. The other wave functions
in (a) have energies in $\left(-2t,0\right)$. Finally, the straight
lines in the innermost circles indicate which sites are joined by
{}``shortcuts''. }
\end{figure}

The above results reflect that there are two quite different types
of states. It is easy to show that the real eigenfunctions of (\ref{eq:hamiltonian})
have the form\begin{equation}
\psi \left(x\right)=B_{l}e^{-iK\left(x-x_{l}\right)}+C_{l}e^{iK\left(x-x_{l}\right)}\label{eq:general-wave-function}\end{equation}
for $x_{l}<x<x_{l+1}$, where $x_{l}$ denotes the $l^{\textrm{th}}$
shortcut terminal (i.e. $x_{l}=i_{s}$ or $j_{s}$ for some $s$),
ordered by ascending site index, and $x_{l+1}-x_{l}>2$. The parameter
$K$ and the corresponding energy $\epsilon _{K}$ can only take one
of the following forms: $K=-i\kappa \Rightarrow \epsilon _{K}=-2t\cosh \kappa $;
$K=k\Rightarrow \epsilon _{K}=-2t\cos k$; $K=\pi -i\kappa \Rightarrow \epsilon _{K}=2t\cosh \kappa $.
Thus in the space between shortcut terminals states with $\left|\epsilon \right|<2t$
are plane waves, with wave number $k=\cos ^{-1}\left(-\epsilon /2t\right)$,
while those with $\pm \epsilon >2t$ decrease or grow exponentially
at the rate $\kappa =\cosh ^{-1}\left(\pm \epsilon /2t\right)$. A
particular case is the ground state of a large chain with a large,
single loop, which has energy $\epsilon =-\sqrt{5}t$ and is exponentially
localised at the two shortcut terminals, with localisation length
$\kappa ^{-1}\sim 2\textrm{ sites}$ \cite{Abrukina-Oksengendler-94}.
The case with more than one shortcut terminal is more complex. A few
wave functions for a particular realisation of our model, with $p\sim 0.01$
shortcuts per site, are shown in Fig.~\ref{cap:wave-functions-polar}.
States with $\epsilon <-2t$ are exponentially localised on some of
the shortcuts. Their energy is very close to $-\sqrt{5}t$, but the
degeneracy is broken by localising on different sets of shortcuts
and by forming bonding and antibonding combinations between them.
This leads to the broadening of the peaks in Fig.~\ref{cap:DOS}.
States with $-2t<\epsilon <0$ resemble plane waves, except that their
amplitude {}``jumps'' at some of the shortcut terminals. 

\section*{Critical behaviour at the lower band edge} The results presented
above indicate a qualitative change of behaviour when $p$ becomes
non-zero. In particular the gradual population of the energy peaks
at $\epsilon =\pm \sqrt{5}t$ is suggestive of a critical point at
$p=0$ in the thermodynamic limit (\ref{eq:thermodynamic-limit}).
This further implies the universal dependence of all observables on
$L/\xi $ in the finite-size scaling regime (\ref{eq:finite-size-scaling}).
The latter has been verified extensively for geometrical features
of the NW and WS networks, such as the shortest path length and the
clustering coefficient \cite{Newman-Watts-99,Barthelemey-Amaral-99,Newman-Watts-99b,Barrat-Weigt-00}.
Here we are concerned with spectral properties and wave functions.
To obtain data above and below the critical system size $\xi =p^{-1}$
we need to consider very large systems. In this case it is best to
use a sparse-matrix numerical algorithm \cite{Farkas-Derenyi-Barabasi-Vicsek-01}.
We employ \cite{ARPACK} which, for real symmetric matrices, is an
implementation of the Lanczos method. For computational economy, given that 
we are now going to examine much larger system sizes, we concentrate on the 
single state at the lower band edge.

\begin{figure}
\includegraphics[  width=0.80\columnwidth,
  keepaspectratio]{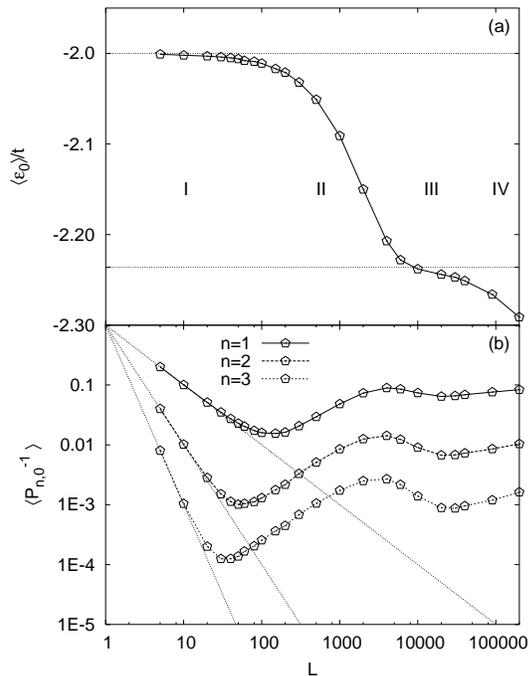}

\caption{\label{cap:lower-band-edge-e-and-IPR}Dependence on the system size
$L$ of (a) the energy and (b) the IPR ($n=1$) and next two higher
moments of the density distribution ($n=2,3$) at the lower band edge
obtained by Lanczos diagonalisation and averaged over 2000 random
realisations. The density of shortcuts is $p=5\times 10^{-4}$. The
dotted lines indicate $\left\langle \epsilon _{0}\right\rangle =-2t$
and $-\sqrt{5}t$ (a) and the analytical results for $p=0$ (b). }
\end{figure}

Fig.~\ref{cap:lower-band-edge-e-and-IPR} shows, for fixed $p$,
the dependence on $L$ of the configurational averages of the ground-state
energy $\epsilon _{0}$ and of the moments of the density distribution
for the ground state wave function, $P_{n,0}^{-1}\equiv \sum _{j}\left|\psi _{0}\left(j\right)\right|^{2\left(n+1\right)}.$
For finite conducting chains at fixed chemical potential, the average of 
$\epsilon_0$ gives a gross idea of how much the energy levels leak out, due 
to the presence of the shortcuts, below the bottom of the original 
conduction band. Our main interest in this quantity, however, is that it 
displays critical behaviour quantitatively consistent with the small-world 
effect (see below). There are four distinct regimes: for $L\ll \xi $ (I) most realisations
do not contain any crossings, so $\left\langle \epsilon _{0}\right\rangle \approx -2t$
and $\left\langle P_{n,0}^{-1}\right\rangle =L^{-n}$. For $L\lesssim \xi $
(II) the system is in a crossover region in which $\left\langle \epsilon _{0}\right\rangle $
decreases and the $\left\langle P_{n,0}^{-1}\right\rangle $ increase
with system size. For $L\gtrsim \xi $ (III) the energy reaches a
plateau and stays close to $\left\langle \epsilon _{0}\right\rangle \approx -\sqrt{5}t$.
In this regime most realisations have one or more shortcuts, but their
terminals are quite far apart. The ground-state wave functions look
like the one in Fig.~\ref{cap:wave-functions-polar}~(a). Interestingly
in this range of values of $L$ the $\left\langle P_{n,0}^{-1}\right\rangle $
decrease again. This is because although \emph{locally} the wave function
is exponentially localised on shortcut terminals, \emph{globally}
the crossings on which it chooses to localise can be quite spread
over the network, and such spread becomes greater as the network grows
in size and more and more crossings become available. Finally for
$L\gg \xi $ (IV) $\left\langle \epsilon _{0}\right\rangle $ decreases
and the $\left\langle P_{n,0}^{-1}\right\rangle $ turn to rise again.
This indicates that in the thermodynamic limit (\ref{eq:thermodynamic-limit})
the lower band edge is at $-\infty $ and contains localised states,
consistent with the Lifshitz tails suggested by Fig.~\ref{cap:DOS}.

\begin{figure}
\includegraphics[  width=0.80\columnwidth,
  keepaspectratio]{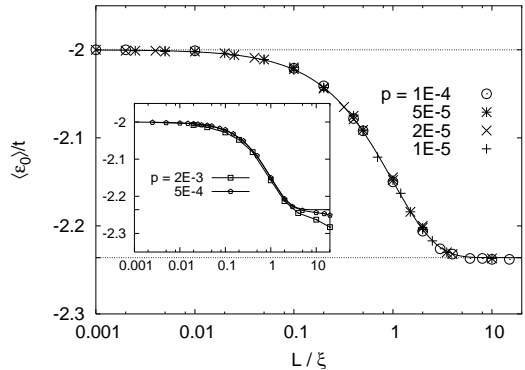}

\caption{\label{cap:Finite-size-scaling-of-ground-state-energy}Finite-size
scaling of the ground state energy, obtained as in Fig.~\ref{cap:lower-band-edge-e-and-IPR}.
The main plot shows the collapse of the data for $p\leq 10^{-4}$
on the curve given by Eq.~(\ref{eq:ground-state-energy-analytical})
for values of the scaling variable from $L/\xi =0.001$ to $L/\xi =20$.
The two dotted lines indicate $\epsilon _{0}=-2t$ and $\epsilon _{0}=-\sqrt{5}t$.
The inset shows data for two higher values of $p$, for comparison.}
\end{figure}

\begin{figure}
\includegraphics[  width=0.75\columnwidth]{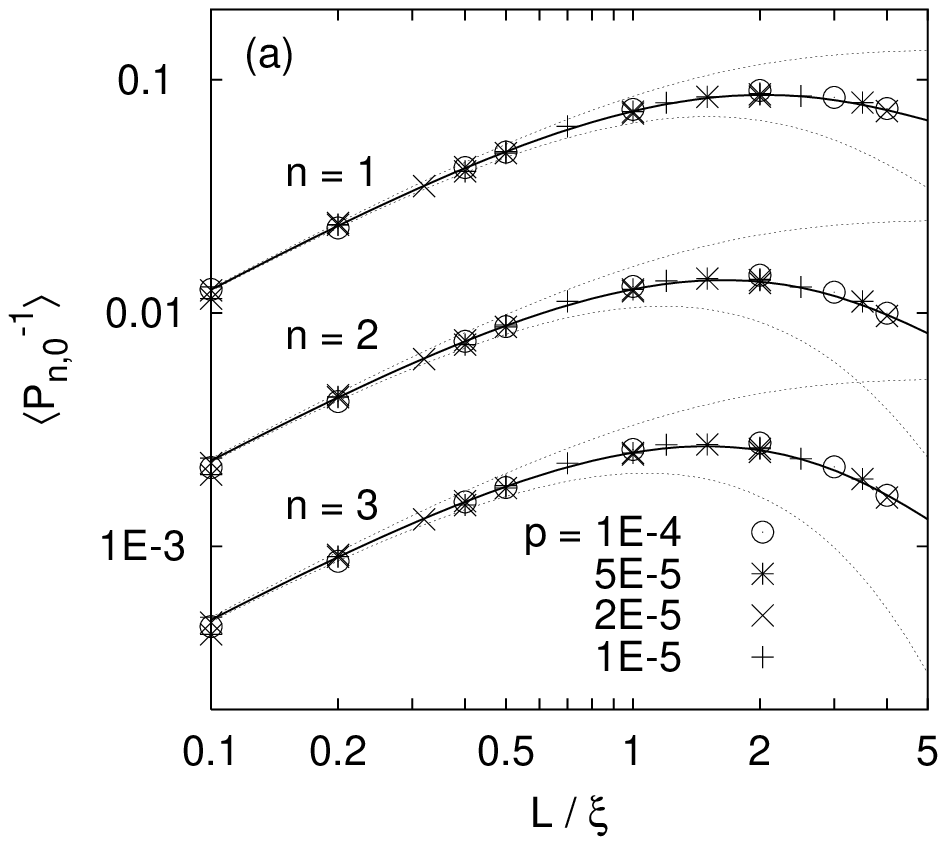}\\
\includegraphics[  width=0.75\columnwidth]{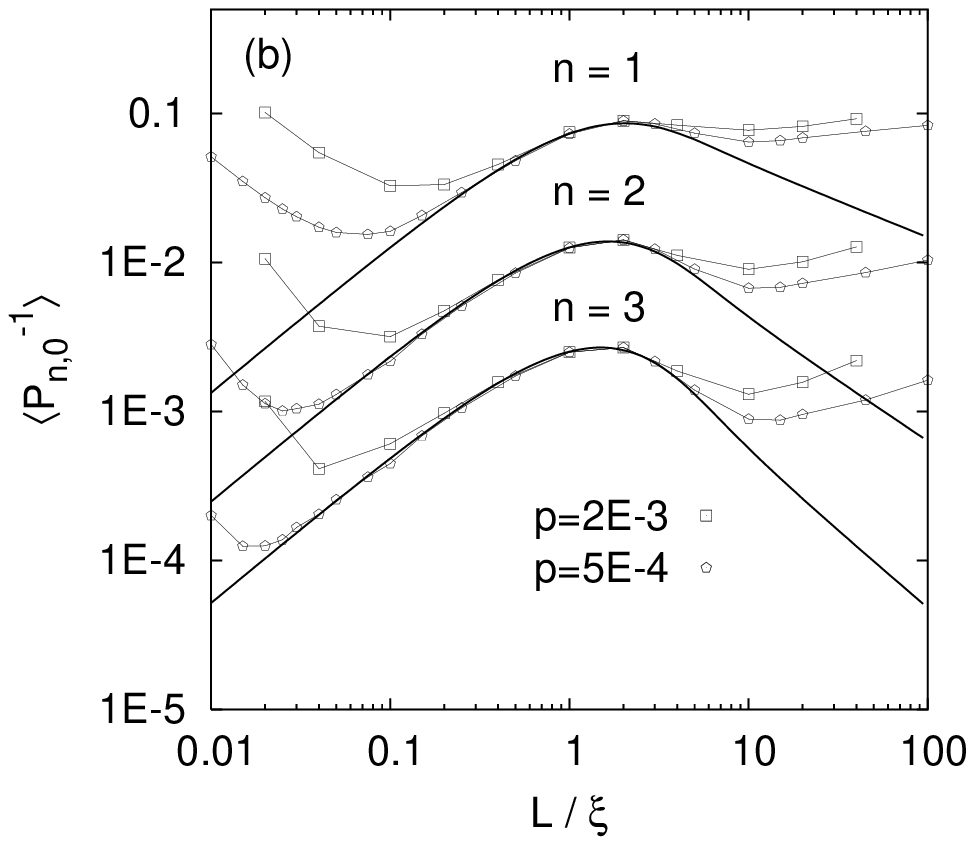}

\caption{\label{cap:Finite-size-scaling-of-the-inverse-participation-ratio}Finite-size
scaling of the IPR ($n=1$) and next two higher moments of the local
density distribution ($n=2,3$) at the lower band edge, obtained as
in Fig.~\ref{cap:lower-band-edge-e-and-IPR}. (a) Shows the collapse
of the data for $p\leq 10^{-4}$ on the curve given by Eq.~(\ref{eq:ground-state-moments-analytical})
for values of the scaling variable from $L/\xi =0.1$ to $L/\xi =5$.
The solid lines correspond to $\alpha =0.48,\, 0.40\, \textrm{and}\, 0.33$
and fit the $n=1,2,3$ data, respectively. The dotted lines correspond
to $\alpha =0$ and $\alpha =1$. In (b) data for two higher values
of $p$ over a wider range of values of $L/\xi $ is displayed for
comparison.}
\end{figure}

Figs.~\ref{cap:Finite-size-scaling-of-ground-state-energy} and \ref{cap:Finite-size-scaling-of-the-inverse-participation-ratio}
show $\left\langle \epsilon _{0}\right\rangle $ and the $\left\langle P_{n,0}^{-1}\right\rangle $
($n=1,2,3$) for very low values of $p$, plotted as functions of
the scaling variable $L/\xi =Lp$. For $p\leq 10^{-4}$ crossings
per site the data collapse to a single curve in the ranges of values
of $L/\xi $ shown. The inset to Fig.~\ref{cap:Finite-size-scaling-of-ground-state-energy}
and Fig.~\ref{cap:Finite-size-scaling-of-the-inverse-participation-ratio}~(b)
show the same data for two higher values of $p$. Evidently the collapse
is not so good. However there is a range of values of $L/\xi $ over
which the data depend only on that variable which becomes wider as
$p$ is lowered. Clearly this widening finite-size scaling regime
corresponds to the regions II and III identified above. 

The defining feature of the finite-size scaling regime is that, although
the number of shortcuts on the network is arbitrary (indeed $\left\langle N\right\rangle =L/\xi $),
all shortcut terminals are well separated. The probability of this
is 
\begin{equation}
P_{\mbox {\small sep}}\approx \exp \left[-4p\left(2Lp-1\right)\kappa ^{-1}\right] \label{eqPsep}
\end{equation}
(taking $N=pL$ and using that the relevant length scale is $4\kappa ^{-1}\ll L$;
see the Appendix).
Evidently in the thermodynamic limit (\ref{eq:thermodynamic-limit})
$P_{\mbox {\small sep}}\to 0$. On the other hand the finite-size
scaling regime is reached when, for a given value of the scaling variable
$Lp$, the density of crossings $p$ is sufficiently low that $P_{\textrm{sep}}\approx 1$.
In this case shortcuts are not close enough to lift the degeneracy
of the ground state and so the only possible energies are $\epsilon _{0}=-2t$
(when there are no shortcuts) and $\epsilon _{0}=-\sqrt{5}t$ (when
there are one or more). Since the probability of the former is $\left(1-\mathcal{P}\right)^{\mathcal{N}}\approx e^{-\mathcal{PN}}$
(for $p\ll L$), 
with $\mathcal{N}$ and $\mathcal{P}$ defined in Eqs.~(\ref{eqN}) and (\ref{eqP}) respectively, 
using $\mathcal{PN}=pL$ we obtain the following
exact analytical expression for the finite-size scaling law giving
the configurational average of the ground-state energy:\begin{equation}
\left\langle \epsilon _{0}\right\rangle =e^{-L/\xi }\left(-2t\right)+\left(1-e^{-L/\xi }\right)\left(-\sqrt{5}t\right).\label{eq:ground-state-energy-analytical}\end{equation}
 This is the curve plotted in Fig.~\ref{cap:Finite-size-scaling-of-ground-state-energy}
alongside the numerical data. Let us now deploy a similar argument
for the moments $\left\langle P_{n,0}^{-1}\right\rangle $. Suppose
that, for a given realisation, having $N$ shortcuts, the wave function
were exponentially localised around a single site $x_{0}$: $\psi _{0}\left(x\right)\sim e^{-\kappa \left|x-x_{0}\right|}$.
Then the moments at the lower band edge would be given by $P_{n,0}^{-1}=\tanh ^{n+1}\kappa /\tanh \kappa \left(n+1\right).$
In actual fact the wave function is localised around $2N'$ sites,
corresponding to the terminals of $N'\leq N$ shortcuts. Since the
sites $x_{t}$ of the terminals are all well separated we can 
derive (see the appendix) 
\begin{equation}
P_{n,0}^{-1}= \frac{\tanh ^{n+1}\kappa}{\left[\tanh \kappa \left(n+1\right)2^{n}N'^{n}\right]}. 
\label{eqPn0}
\end{equation}
Let us assume that $N'$ depends only on $N$ and postulate a power
law: $N'=N^{\alpha }$. Then using the Poisson distribution we obtain\begin{equation}
\left\langle P_{n,0}^{-1}\right\rangle =\sum _{N=1}^{\infty }\frac{e^{-L/\xi }\left(L/\xi \right)^{N}}{N!}\frac{\tanh ^{n+1}\kappa }{\tanh \kappa \left(n+1\right)}\frac{1}{2^{n}N^{\alpha n}}\label{eq:ground-state-moments-analytical}\end{equation}
 which gives the finite-size scaling of the moments $\left\langle P_{n,0}^{-1}\right\rangle $.
It is important to note however that this expression, unlike (\ref{eq:ground-state-energy-analytical}),
contains an adjustable parameter $\alpha $. Depending on its value
it can describe very different behaviours. Indeed for $\alpha =1$
(i.e. requiring that the wave function is exponentially localised
always on the same fraction of the shortcuts) we have\begin{equation}
\left\langle P_{1,0}^{-1}\right\rangle =\frac{\tanh ^{n+1}\kappa }{2\tanh \kappa \left(n+1\right)}\left\{ \textrm{Ei}\left(L/\xi \right)-\ln \left(L/\xi \right)-\gamma \right\} ,\end{equation}
where $\gamma $ is Euler's constant and $\textrm{Ei}\left(x\right)$
is the exponential integral. For large $L/\xi $, we have $\left\langle P_{1,0}^{-1}\right\rangle \sim \left(L/\xi \right)^{-1}$
(more generally $\left\langle P_{n,0}^{-1}\right\rangle \sim \left(L/\xi \right)^{-n}$)
corresponding to a state which, although exponentially localised at
the \emph{local} level, is extended \emph{globally} over the whole
network. Conversly for $\alpha =0$ (i.e. the wave function is exponentially
localised always on a single shortcut) we obtain $\left\langle P_{n,0}^{-1}\right\rangle \approx \textrm{constant}$
for large $L/\xi $, i.e. the wave function is localised in the true
sense of the word. The dashed lines in Fig.~\ref{cap:Finite-size-scaling-of-the-inverse-participation-ratio}~(a)
represent these two extreme cases. On the other hand the data are
fitted quite accurately by using three different intermediate values
of $\alpha $ for $n=1,2,3$, respectively. This is represented by
the solid curves in Fig.~\ref{cap:Finite-size-scaling-of-the-inverse-participation-ratio}.
It suggests that the \emph{global} structure of the wave function
at the lower band edge is \emph{multifractal} in the finite-size scaling
regime.

\section*{Conclusion} In summary we have studied a simple model of conducting
polymers in which chain crossings act as shortcuts for the electrons,
so they move in a small-world topology. Our results are based on numerical
diagonalisation (full spectrum for systems of size $L\sim 10^{3}$
and Lanczos for $L\sim 10^{5}$) and some simple analytical arguments.
We have seen that a small density of shortcuts $p$ leads to qualitative,
but continuous changes in the DOS and wave functions. The former take
the form of new peaks appearing at well defined positions $\epsilon =\pm \sqrt{5}t$
outside the initial conduction band, $\left|\epsilon \right|<2t$.
At these peaks, the wave functions have a non-trivial structure. We
have investigated in detail the lower band edge, confirming through
finite-size scaling that there is a critical point at $p=0$. The
critical exponent of the correlation length is consistent with criticality
due to the small-world effect. We have derived analytical expressions
for the scaling laws of the energy and the IPR and higher moments
of the density distribution. The latter expressions contain an adjustable
parameter $\alpha $ describing the global structure of the wave function,
which we find to be multifractal. 

We end by making two additional remarks. Firstly, the critical point
described here is quite distinct from the dimensionality-driven Anderson
transition considered before \cite{Prigodin-Efetov-93,Zambetakiy-Evangelou-Economou-96,Dupuis-97,Zhu-Xiong-00,Zhu-Xiong-01}.
The latter occurs at a finite critical density of crossings $p_{c}\left(W\right)>0$,
which depends on the amount of site-energy disorder $W$. Since real
highly conducting polymers are intrinsically disordered \cite{Heeger-Kivelson-Schrieffer-Su-88}
the small-world critical point at $p=0$ is not directly accessible
to experiments. Nevertheless it may make a significant contribution
to conductance fluctuations in the metallic state. Secondly, the multifractality
of the wave functions suggests that the adjacency matrices of small-world
networks might be describable by the power law banded matrix ensemble
\cite{Kravtsov-Muttalib-97} of random matrix theory. This intriguing
possibility remains largely unexplored %
\footnote{We acknowledge V. E. Kravtsov and B. N. Narozhny for this suggestion.%
}.

\begin{acknowledgments}
We thank P.R.A. Campos, K. Capelle, E.C. Carter, J.M.F. Gunn, C. Hooley,
V.E. Kravtsov, I.V. Lerner, B.N. Narozhny, L.N. Oliveira, A.J. Schofield,
R.A. Smith and I.V. Yurkevich for useful discussions. JQ thanks the
Abdus Salam ICTP for hospitality during the preparation of part of
this manuscript. We acknowledge financial support from CAPES (VLC),
FAPESP and The Leverhulme Trust (JQ). In addition, JQ gratefully acknowledges an Atlas
fellowship granted by CCLRC in association with St. Catherine's College in the
Univeristy of Oxford.
\end{acknowledgments}

\appendix

\section*{Appendix}

In this Appendix we provide some simple derivations omitted in the main text. 
Firstly, let us prove that the
probability $P_{\mbox {\small sep}}$ of all the $2N$ 
shortcut terminals being well separated from each other is given by Eq.~(\ref{eqPsep}). 
Each terminal is found at a random location along the chain. Let $x_t$, with
$t=1,2,...,2N,$ be the site indices of those locations. 
As noted in the main text, the ground-state wave function decays exponentially 
as $e^{-\kappa|x - x_t|}$ in the space between terminals. After selecting the
location of the first 
terminal, the probability for at least one of the $2N - 1$ remaining terminals 
to be within $2\kappa^{-1}$ of $x_1$ (the factor of two is arbitrary and does not affect
the argument) is 
$(4\kappa^{-1} / L) (2N - 1)$. Thus the probability that the first terminal is 
separated form all the other terminals (in the sense that it is within two decay
lengths of any one of them) is
\begin{equation}
P_1 =  1 - \frac{4\kappa^{-1}}{L}(2N - 1).
\end{equation}
Now consider the second terminal, located at $x_2$. The probability of the remaining
$2N - 2$ terminals being well separated from this one, in the same sense as
above, is $1 - (4\kappa^{-1}/L)(2N - 2)$. So the probability of both the first and 
second terminals being well separated from all the others is
\begin{equation}
P_2 = \left[1 - \frac{4\kappa^{-1}}{L}\left(2N - 1\right)\right]
\left[1 - \frac{4\kappa^{-1}}{L}\left(2N - 2\right)\right].
\end{equation}
At this point we can already see that the probability of all the terminals being
well separated from each other is given by 
\begin{equation}
P_{\mbox {\small sep}} = P_{2N-1} = 
\prod_{j=1}^{2N-1}\left[1 - \frac{4\kappa^{-1}}{L}\left(2N - j\right)\right] 
\end{equation}
whence, taking logarithms,
\begin{equation}
\log(P_{\mbox {\small sep}}) \approx 
-\frac{4\kappa^{-1}}{L} \sum_{j=1}^{2N-1} 2N-j = -\frac{4\kappa^{-1}N}{L}(2N-1).
\end{equation}
Considering that the averaged number of shortcuts is $pL$, 
Eq.(\ref{eqPsep}) follows.

Let us now prove Eq.~(\ref{eqPn0}) for the moments of the density 
distribution of the ground-state wave function. With $N$ well separated 
shortcuts, let us suppose the ground-state wave function has peaks at $2N'$  
shortcut terminals ($N' \le N$), decaying exponentially from each maximum again as 
$e^{-\kappa|x-x_t|}$. The moment $P^{-1}_{n,0}$ is given by
\begin{eqnarray}
P^{-1}_{n,0} &=& \sum_{j=1}^L |\psi_0(j)|^{2n+2} \nonumber \\
&\approx& |A|^{2n+2} \sum_{t=1}^{2N'} 
\left[ \sum_{j=-\infty}^{\infty} (e^{-\kappa |j|})^{2n+2} \right] \nonumber \\
&=& |A|^{2n+2} \frac{2N'}{\tanh\left[\left(n+1\right)\kappa\right]}, \label{eqap1}
\end{eqnarray}
where $A$ is a normalization constant. The latter can be found from
$P^{-1}_{0,0}=1$. It is
\begin{equation}
A^2 = \frac{\tanh(\kappa)}{2N'},
\end{equation}
which substituted in Eq.~(\ref{eqap1}) yields Eq.~(\ref{eqPn0}).

\bibliographystyle{apsrev}

\begin{thebibliography}{26}
\expandafter\ifx\csname natexlab\endcsname\relax\def\natexlab#1{#1}\fi
\expandafter\ifx\csname bibnamefont\endcsname\relax
  \def\bibnamefont#1{#1}\fi
\expandafter\ifx\csname bibfnamefont\endcsname\relax
  \def\bibfnamefont#1{#1}\fi
\expandafter\ifx\csname citenamefont\endcsname\relax
  \def\citenamefont#1{#1}\fi
\expandafter\ifx\csname url\endcsname\relax
  \def\url#1{\texttt{#1}}\fi
\expandafter\ifx\csname urlprefix\endcsname\relax\def\urlprefix{URL }\fi
\providecommand{\bibinfo}[2]{#2}
\providecommand{\eprint}[2][]{\url{#2}}

\bibitem[{\citenamefont{Albert and Barab{\'a}si}(2002)}]{Albert-Barabasi-02}
\bibinfo{author}{\bibfnamefont{R.}~\bibnamefont{Albert}} \bibnamefont{and}
  \bibinfo{author}{\bibfnamefont{A.-L.} \bibnamefont{Barab{\'a}si}},
  \bibinfo{journal}{Rev. Mod. Phys.} \textbf{\bibinfo{volume}{74}},
  \bibinfo{pages}{47} (\bibinfo{year}{2002}).

\bibitem[{\citenamefont{Watts and Strogatz}(1998)}]{Watts-Strogatz-98}
\bibinfo{author}{\bibfnamefont{D.~J.} \bibnamefont{Watts}} \bibnamefont{and}
  \bibinfo{author}{\bibfnamefont{S.~H.} \bibnamefont{Strogatz}},
  \bibinfo{journal}{Nature} \textbf{\bibinfo{volume}{393}},
  \bibinfo{pages}{440} (\bibinfo{year}{1998}).

\bibitem[{\citenamefont{Newman and
  Watts}(1999{\natexlab{a}})}]{Newman-Watts-99}
\bibinfo{author}{\bibfnamefont{M.}~\bibnamefont{Newman}} \bibnamefont{and}
  \bibinfo{author}{\bibfnamefont{D.}~\bibnamefont{Watts}},
  \bibinfo{journal}{Phys. Lett. A} \textbf{\bibinfo{volume}{263}},
  \bibinfo{pages}{341} (\bibinfo{year}{1999}{\natexlab{a}}).

\bibitem[{\citenamefont{Newman and
  Watts}(1999{\natexlab{b}})}]{Newman-Watts-99b}
\bibinfo{author}{\bibfnamefont{M.}~\bibnamefont{Newman}} \bibnamefont{and}
  \bibinfo{author}{\bibfnamefont{D.}~\bibnamefont{Watts}},
  \bibinfo{journal}{Phys. Rev. E} \textbf{\bibinfo{volume}{60}},
  \bibinfo{pages}{7332} (\bibinfo{year}{1999}{\natexlab{b}}).

\bibitem[{\citenamefont{Bath{\'e}l{\'e}my and
  Amaral}(1999)}]{Barthelemey-Amaral-99}
\bibinfo{author}{\bibfnamefont{M.}~\bibnamefont{Bath{\'e}l{\'e}my}}
  \bibnamefont{and} \bibinfo{author}{\bibfnamefont{L.~A.~N.}
  \bibnamefont{Amaral}}, \bibinfo{journal}{Phys. Rev. Lett.}
  \textbf{\bibinfo{volume}{82}}, \bibinfo{pages}{3180} (\bibinfo{year}{1999}),
  \bibinfo{note}{{\emph{erratum}}: {\bf 82}, 5180 (1999).}

\bibitem[{\citenamefont{Barrat and Weigt}(2000)}]{Barrat-Weigt-00}
\bibinfo{author}{\bibfnamefont{A.}~\bibnamefont{Barrat}} \bibnamefont{and}
  \bibinfo{author}{\bibfnamefont{M.}~\bibnamefont{Weigt}},
  \bibinfo{journal}{Eur. Phys. J. B} \textbf{\bibinfo{volume}{13}},
  \bibinfo{pages}{547} (\bibinfo{year}{2000}).

\bibitem[{\citenamefont{Prigodin and Efetov}(1993)}]{Prigodin-Efetov-93}
\bibinfo{author}{\bibfnamefont{V.}~\bibnamefont{Prigodin}} \bibnamefont{and}
  \bibinfo{author}{\bibfnamefont{K.}~\bibnamefont{Efetov}},
  \bibinfo{journal}{Phys. Rev. Lett.} \textbf{\bibinfo{volume}{70}},
  \bibinfo{pages}{2932} (\bibinfo{year}{1993}).

\bibitem[{\citenamefont{Zambetakiy et~al.}(1996)\citenamefont{Zambetakiy,
  Evangelou, and Economou}}]{Zambetakiy-Evangelou-Economou-96}
\bibinfo{author}{\bibfnamefont{I.}~\bibnamefont{Zambetakiy}},
  \bibinfo{author}{\bibfnamefont{S.~N.} \bibnamefont{Evangelou}},
  \bibnamefont{and} \bibinfo{author}{\bibfnamefont{E.~N.}
  \bibnamefont{Economou}}, \bibinfo{journal}{J. Phys. : Condens. Matter}
  \textbf{\bibinfo{volume}{8}}, \bibinfo{pages}{L605} (\bibinfo{year}{1996}).

\bibitem[{\citenamefont{Dupuis}(1997)}]{Dupuis-97}
\bibinfo{author}{\bibfnamefont{N.}~\bibnamefont{Dupuis}},
  \bibinfo{journal}{Phys. Rev. B} \textbf{\bibinfo{volume}{56}},
  \bibinfo{pages}{3086} (\bibinfo{year}{1997}).

\bibitem[{\citenamefont{Zhu and Xiong}(2000)}]{Zhu-Xiong-00}
\bibinfo{author}{\bibfnamefont{C.-P.} \bibnamefont{Zhu}} \bibnamefont{and}
  \bibinfo{author}{\bibfnamefont{S.-J.} \bibnamefont{Xiong}},
  \bibinfo{journal}{Phys. Rev. B} \textbf{\bibinfo{volume}{62}},
  \bibinfo{pages}{14780} (\bibinfo{year}{2000}).

\bibitem[{\citenamefont{Zhu and Xiong}(2001)}]{Zhu-Xiong-01}
\bibinfo{author}{\bibfnamefont{C.-P.} \bibnamefont{Zhu}} \bibnamefont{and}
  \bibinfo{author}{\bibfnamefont{S.-J.} \bibnamefont{Xiong}},
  \bibinfo{journal}{Phys. Rev. B} \textbf{\bibinfo{volume}{63}},
  \bibinfo{pages}{193405} (\bibinfo{year}{2001}).

\bibitem[{\citenamefont{Xiong and Evangelou}(1995)}]{Xiong-Evangelou-95}
\bibinfo{author}{\bibfnamefont{S.-J.} \bibnamefont{Xiong}} \bibnamefont{and}
  \bibinfo{author}{\bibfnamefont{S.}~\bibnamefont{Evangelou}},
  \bibinfo{journal}{Phys. Rev. B} \textbf{\bibinfo{volume}{52}},
  \bibinfo{pages}{R13079} (\bibinfo{year}{1995}).

\bibitem[{\citenamefont{Xiong et~al.}(1996)\citenamefont{Xiong, Chen, and
  Evangelou}}]{Xiong-Chen-Evangelou-96}
\bibinfo{author}{\bibfnamefont{S.-J.} \bibnamefont{Xiong}},
  \bibinfo{author}{\bibfnamefont{Y.}~\bibnamefont{Chen}}, \bibnamefont{and}
  \bibinfo{author}{\bibfnamefont{S.~N.} \bibnamefont{Evangelou}},
  \bibinfo{journal}{Phys. Rev. Lett.} \textbf{\bibinfo{volume}{77}},
  \bibinfo{pages}{4414} (\bibinfo{year}{1996}).

\bibitem[{\citenamefont{Farkas et~al.}(2001)\citenamefont{Farkas, Derenyi,
  Barabasi, and Vicsek}}]{Farkas-Derenyi-Barabasi-Vicsek-01}
\bibinfo{author}{\bibfnamefont{I.~J.} \bibnamefont{Farkas}},
  \bibinfo{author}{\bibfnamefont{I.}~\bibnamefont{Derenyi}},
  \bibinfo{author}{\bibfnamefont{A.~L.} \bibnamefont{Barabasi}},
  \bibnamefont{and} \bibinfo{author}{\bibfnamefont{T.}~\bibnamefont{Vicsek}},
  \bibinfo{journal}{Phys. Rev. E} \textbf{\bibinfo{volume}{64}},
  \bibinfo{pages}{026704} (\bibinfo{year}{2001}).

\bibitem[{\citenamefont{Kim et~al.}(2003{\natexlab{a}})\citenamefont{Kim, Hong,
  and Choi}}]{Kim-Hong-Choi-03}
\bibinfo{author}{\bibfnamefont{B.~J.} \bibnamefont{Kim}},
  \bibinfo{author}{\bibfnamefont{H.}~\bibnamefont{Hong}}, \bibnamefont{and}
  \bibinfo{author}{\bibfnamefont{M.~Y.} \bibnamefont{Choi}},
  \bibinfo{journal}{Phys. Rev. B} \textbf{\bibinfo{volume}{68}},
  \bibinfo{pages}{014304} (\bibinfo{year}{2003}{\natexlab{a}}).

\bibitem[{\citenamefont{Kramer and MacKinnon}(1993)}]{Kramer-MacKinnon-93}
\bibinfo{author}{\bibfnamefont{B.}~\bibnamefont{Kramer}} \bibnamefont{and}
  \bibinfo{author}{\bibfnamefont{A.}~\bibnamefont{MacKinnon}},
  \bibinfo{journal}{Rep. Prog. Phys.} \textbf{\bibinfo{volume}{56}}
  (\bibinfo{year}{1993}).

\bibitem[{\citenamefont{Abrukina and
  Oxsengendler}(1994)}]{Abrukina-Oksengendler-94}
\bibinfo{author}{\bibfnamefont{Y.~M.} \bibnamefont{Abrukina}} \bibnamefont{and}
  \bibinfo{author}{\bibfnamefont{B.}~\bibnamefont{Oxsengendler}},
  \textbf{\bibinfo{volume}{60}}, \bibinfo{pages}{270} (\bibinfo{year}{1994}).

\bibitem[{\citenamefont{Lehoucq et~al.}()\citenamefont{Lehoucq, Sorensen, and
  Yang}}]{ARPACK}
\bibinfo{author}{\bibfnamefont{R.}~\bibnamefont{Lehoucq}},
  \bibinfo{author}{\bibfnamefont{D.}~\bibnamefont{Sorensen}}, \bibnamefont{and}
  \bibinfo{author}{\bibfnamefont{C.}~\bibnamefont{Yang}},
  \bibinfo{note}{{ARPACK} library. http://www.caam.rice.edu/software/ARPACK}.

\bibitem[{\citenamefont{Heeger et~al.}(1988)\citenamefont{Heeger, Kivelson,
  Schrieffer, and Su}}]{Heeger-Kivelson-Schrieffer-Su-88}
\bibinfo{author}{\bibfnamefont{A.~J.} \bibnamefont{Heeger}},
  \bibinfo{author}{\bibfnamefont{S.}~\bibnamefont{Kivelson}},
  \bibinfo{author}{\bibfnamefont{J.~R.} \bibnamefont{Schrieffer}},
  \bibnamefont{and} \bibinfo{author}{\bibfnamefont{W.~P.} \bibnamefont{Su}},
  \bibinfo{journal}{Rev. Mod. Phys.} \textbf{\bibinfo{volume}{60}},
  \bibinfo{pages}{781} (\bibinfo{year}{1988}).

\bibitem[{\citenamefont{Kravtsov and Muttalib}(1997)}]{Kravtsov-Muttalib-97}
\bibinfo{author}{\bibfnamefont{V.~E.} \bibnamefont{Kravtsov}} \bibnamefont{and}
  \bibinfo{author}{\bibfnamefont{K.~A.} \bibnamefont{Muttalib}},
  \bibinfo{journal}{Phys. Rev. Lett.} \textbf{\bibinfo{volume}{79}},
  \bibinfo{pages}{1913} (\bibinfo{year}{1997}).

\bibitem[{\citenamefont{Luque and Miramontes}()}]{Luque-Miramontes-02}
\bibinfo{author}{\bibfnamefont{B.}~\bibnamefont{Luque}} \bibnamefont{and}
  \bibinfo{author}{\bibfnamefont{O.}~\bibnamefont{Miramontes}},
  \bibinfo{note}{cond-mat/0211383}.

\bibitem[{\citenamefont{Metzler et~al.}(2002)}]{Metzler-et-al-02}
\bibinfo{author}{\bibfnamefont{R.}~\bibnamefont{Metzler}} \bibnamefont{et~al.},
  \bibinfo{journal}{Phys. Rev. Lett.} \textbf{\bibinfo{volume}{88}},
  \bibinfo{pages}{188101} (\bibinfo{year}{2002}).

\bibitem[{\citenamefont{Bao et~al.}(2003)\citenamefont{Bao, Lee, and
  Quake}}]{Bao-Lee-Quake-03}
\bibinfo{author}{\bibfnamefont{X.~R.} \bibnamefont{Bao}},
  \bibinfo{author}{\bibfnamefont{H.~J.} \bibnamefont{Lee}}, \bibnamefont{and}
  \bibinfo{author}{\bibfnamefont{S.~R.} \bibnamefont{Quake}},
  \bibinfo{journal}{Phys. Rev. Lett.} \textbf{\bibinfo{volume}{91}},
  \bibinfo{pages}{265506} (\bibinfo{year}{2003}).

\bibitem[{\citenamefont{Monasson}(1999)}]{Monasson-99}
\bibinfo{author}{\bibfnamefont{R.}~\bibnamefont{Monasson}},
  \bibinfo{journal}{Eur. Phys. J. B} \textbf{\bibinfo{volume}{12}},
  \bibinfo{pages}{555} (\bibinfo{year}{1999}).

\bibitem[{\citenamefont{Kim et~al.}(2003{\natexlab{b}})\citenamefont{Kim, Hong,
  and Choi}}]{Kim-Hong-Choi-03b}
\bibinfo{author}{\bibfnamefont{B.~J.} \bibnamefont{Kim}},
  \bibinfo{author}{\bibfnamefont{H.}~\bibnamefont{Hong}}, \bibnamefont{and}
  \bibinfo{author}{\bibfnamefont{M.~Y.} \bibnamefont{Choi}},
  \bibinfo{journal}{J. Phys. A: Math. Gen.} \textbf{\bibinfo{volume}{36}},
  \bibinfo{pages}{6329} (\bibinfo{year}{2003}{\natexlab{b}}).

\bibitem[{\citenamefont{Eaton et~al.}()}]{Octave}
\bibinfo{author}{\bibfnamefont{J.~W.} \bibnamefont{Eaton}}
  \bibnamefont{et~al.}, \bibinfo{note}{{Octave} package.
  http://www.octave.org}.

\end{thebibliography}

\end{document}